\documentclass[aps,prd,onecolumn,superscriptaddress,nofootinbib,10pt]{revtex4-1}
\usepackage[colorlinks,linkcolor=blue,anchorcolor=blue,citecolor=blue,urlcolor=blue]{hyperref}
\usepackage{graphicx,subfigure}
\usepackage[figuresright]{rotating}
\usepackage{amsmath,amsfonts,amssymb,bm}
\usepackage{array,enumitem,multirow}
\usepackage{acronym}
\usepackage{makecell}
\newcommand{\be}{\begin{equation}}
\newcommand{\ee}{\end{equation}}
\newcommand{\bea}{\begin{eqnarray}}
\newcommand{\eea}{\end{eqnarray}}
\newcommand{\nn}{\nonumber}

\newcommand{\mSun}{{\rm~M_\odot}}
\newcommand{\cO}{{\cal O}}
\newcommand{\TQC}{MOE Key Laboratory of TianQin Mission, TianQin Research Center for Gravitational Physics \&  School of Physics and Astronomy, Frontiers Science Center for TianQin, Gravitational Wave Research Center of CNSA, Sun Yat-sen University (Zhuhai Campus), Zhuhai 519082, China.}

\newcommand{\mnras}{Monthly Notices of the Royal Astronomical Society}
\newcommand{\apjl}{the Astrophysical Journal Letters}

\newacro{GR}{General Relativity}
\newacro{GW}{gravitational wave}
\newacro{CMB}{Cosmic Microwave Background}
\newacro{LSS}{last-scattering surface}
\newacro{FLRW}{Friedmann-Lema\^{\i}tre-Robertson-Walker}
\newacro{BCS}{Bounded Cosmic Space}
\newacro{CB}{Cosmic Boundary}
\newacro{SBH}{stellar-mass black hole}
\newacro{MBH}{massive black hole}
\newacro{PBH}{primordial black hole}
\newacro{SBHB}{stellar-mass black hole binary}
\newacro{MBHB}{massive black hole binary}
\newacro{IMRI}{intermediate mass ratio inspiral}
\newacro{LVK}{the LIGO-Virgo-Kagra collaboration}
\newacro{CE}{Cosmic Explorer}
\newacro{ET}{Einstein Telescope}
\newacro{LISA}{Laser Interferometer Space Antenna}
\newacro{SNR}{signal-to-noise ratio}
\newacro{FIM}{Fisher Information Matrix}
\newacro{MCMC}{Markov Chain Monte Carlo}
\begin{document}

\title{Gravitational waves and cosmic boundary}

\author{Changfu Shi}
\author{Xinyi Che}
\author{Zeyu Huang}
\author{Yi-Ming Hu}
\author{Jianwei Mei}
\email{Email: meijw@sysu.edu.cn}
\affiliation{\TQC}

\date{\today}

\begin{abstract}
Space-based gravitational wave detectors have the capability to detect signals from very high redshifts.
It is interesting to know if such capability can be used to study the global structure of the cosmic space.
In this paper, we focus on one particular question:
if there exists a reflective cosmic boundary at the high redshift ($z>15$), is it possible to find it?
We find that, with the current level of technology:
1) gravitational waves appear to be the only means with which that signatures from the cosmic boundary can possibly be detected;
2) a large variety of black holes, with masses roughly in the range $\cO(10^3\sim10^6)\mSun$, can be used for the task;
3) in the presumably rare but physically possible case that two merger events from the growth history of a massive black hole are detected coincidentally, a detector network like TianQin+LISA is essential in help improving the chance to determine the orientation of the cosmic boundary;
4) the possibility to prove or disprove the presence of the cosmic boundary largely depends on how likely one can detect multiple pairs of coincident gravitational wave events.
\end{abstract}

\maketitle

\section{Introduction}
\label{sec:intro}

What is the size and topology of the three dimensional cosmic space? This is one basic question that still lacks an answer.
The observable portion of the cosmic space is consistent with being flat \cite{Planck:2018vyg,Lahav:2024npe}.
For a three dimensional flat space, there are 18 possible topologies, with the simplest being the infinitely large Euclidean space $\mathbb{R}^3$, and the simplest nontrivial case being the 3-torus, $T^3$, made from a rectangular prism with each pair of its opposite faces simply identified.
All the 17 nontrivial topologies involve compact directions supporting closed geodesics that cannot be shrunk to a point continuously, and they all give raise to a multi-connected universe without a boundary.
For a detailed description of Euclidean topologies relevant for cosmology, we refer to \cite{COMPACT:2023rkp}.

Nontrivial cosmic topology can have nontrivial impact on the \ac{CMB}.
\ac{CMB} photons we see today have been emitted from the \ac{LSS} at recombination.
The \ac{LSS} bounds the portion of the cosmic space that has been directly observed.
One way to search for nontrivial topology is to look for the corresponding modification of the \ac{CMB} spectrum \cite{Lachieze-Rey:1995qrb,Levin:2001fg}.
This method is model dependent and faces the difficulty of having unlimited possibilities on the choice of topologies and their parameters.
Another method is to look for matched circles on the \ac{CMB} sky \cite{Cornish:1997ab}.
If the cosmic space has a nontrivial topology, such as $T^3$, an observer would see (assuming unlimited light power and waiting time) infinitely many copies of herself making up a three dimensional grid.
Each grid point has a copy of the \ac{LSS} to itself.
If there is a compact direction smaller than the diameter of the \ac{LSS}, then the \acp{LSS} of two nearby grid points will intersect each other.
Since all grid points are in fact one of the same, the circle where the two nearby \acp{LSS} intersect will be seen by the observer as two circles located in opposite directions of the sky but with matched temperature variations.
So far all matched circle searches have yielded null results \cite{deOliveira-Costa:2003utu,Cornish:2003db,ShapiroKey:2006hm,Mota:2010jb,Bielewicz:2010bh,Bielewicz:2011jz,Vaudrevange:2012da,Aurich:2013fwa,Planck:2013okc,Planck:2015gmu}.
Since different topologies and choices of parameters predict different patterns of matched circles, such null results mean differently for different topologies, with many still allowed to have a compact direction smaller than the diameter of the \ac{LSS} \cite{COMPACT:2022nsu,COMPACT:2022gbl}.
However, the null results do exclude all closed and Earth (observer) passing geodesics that are topologically nontrivial and are shorter than the diameter of the \ac{LSS}.
So one is not expected to detect multiple images from a same astrophysical object due to nontrivial topology.
Of course, this may change in the future when the size of the \ac{LSS} has become large enough.
This can be tested by observing distant galaxies, see \cite{Fang_1983,1987ApJ...322L...5F,Roukema:1996cu,Lehoucq:1996qe,Luminet:1999qh,Fujii:2011ga,Fujii:2013xsa,Roukema:2013jxc,Luminet:2016bqv,2003MNRAS.342L...9W} and references therein.

The null results on matched circles also mean that there will be no observationally nontrivial signatures for individual \ac{GW} events.
But a flat cosmic space can have nontrivial global structures other than the topologies mentioned above.
For example, all the 18 topologies mentioned above describe a cosmic space {\it without} boundaries, but one may as well consider a cosmic space {\it with} boundaries.
The simplest example is the simply connected Euclidean space bounded from all directions.
Such cosmic space is topologically equivalent to a 3-disk, $D^3$.
For later convenience, we will refer to such model of the cosmic space the \ac{BCS} and its boundaries the \ac{CB}.
Some theoretical motivation for the \ac{BCS} will be given in the next section.

There is no reason to believe that the size of the \ac{BCS} is tied to that of the \ac{LSS}.
Instead, the former could be much larger than the latter.
In this case, the \ac{BCS} will not be relevant for observation unless part of the \ac{CB} happens to be enclosed by the \ac{LSS} (see Fig. \ref{fig:CB} for an illustration).
To be consistent with the \ac{CMB} observations, the \ac{CB} should be reflective.
In this case, the \ac{CB} acts like a mirror and evidence for the \ac{CB} can be obtained if one can identify an astrophysical object together with its image in the \ac{CB}.
If the \ac{CB} is close enough, one may try to observe it with electromagnetic telescopes.
For example, the current most distant known galaxy has been found by JWST at a redshift $z\approx14.3$ \cite{2024arXiv240518485C}.
If the \ac{CB} is further away, one will need a method with even deeper spatial reach.
Fortunately, the breakthrough in \ac{GW} detection has greatly enhanced our capability to detect physics from the high redshifts.
For example, \ac{GW} detectors planned for the 2030s will not have much difficulty in reaching redshifts $z>15$ \cite{eLISA:2013xep,LISA:2017pwj,Luo:2015ght,Hu:2017yoc,Wang:2019ryf,TianQin:2020hid,Torres-Orjuela:2023hfd,Hu:2017mde,Evans:2021gyd,Maggiore:2019uih,Bailes:2021tot,Branchesi:2023mws}.

In this paper, we focus on using \ac{GW} signals from \ac{MBH} systems to explore the possible presence of a reflective \ac{CB}.
Revealing the formation and growth history of \acp{MBH} is one of the most important scientific objectives of the upcoming space-based \ac{GW} detectors \cite{LISA:2022yao,Wang:2019ryf}.
It is likely that a \ac{MBH} can experience many mergers throughout its growth history \cite{Dayal:2018gwg,Pacucci:2020orw,Piana:2020rwb,Liu:2024kig}.
If a \ac{MBH} is located close enough to the \ac{CB} and has merger events at the right times, then a detector might be able to detect two merger events from the history of the same \ac{MBH}, with the earlier signal taking the longer route that involves a reflection from the \ac{CB}.
Such pairs of coincident signals could be rare but are physically possible.
If detected and confirmed, such signals will not only provide evidence for the existence of the \ac{CB}, but will also offer valuable information on the growth history of the \ac{MBH}.
The purpose of this paper is to study such detection scenario in detail and to find the relevant parameter space for the \acp{MBH}.

The rest of the paper is organized as following.
In section \ref{sec:cb}, we present some theoretical motivation for the \ac{BCS} and discuss the possible properties of the \ac{CB}.
In section \ref{sec:scheme}, we study the detection scenario and the relevant parameter space in detail.
In section \ref{sec:sum}, we conclude with a short summary and discussion.

\section{The Bounded Cosmic Space and the cosmic boundary}
\label{sec:cb}

The idea for a finite and bounded cosmic space can naturally arise from the perspective of emergent gravity (see \cite{Hu:2009jd,Sindoni:2011ej,Carlip:2012wa,Linnemann:2017hdo} and references therein).
In emergent gravity, various aspects of gravity, such as the spacetime manifold, the spacetime metric, and the dynamics of the spacetime metric, can all be emergent from some more fundamental construction, such as some fundamental particles (e.g., \cite{Hu:2009jd,Sindoni:2011ej}).
Regardless of their detailed properties, a system of interacting particles are expected to collectively behave like a fluid at the macroscopic scale.
This observation has been used to help understand the fluid/gravity correspondence \cite{Bhattacharyya:2007vjd,Rangamani:2009xk,Hubeny:2011hd} based on the AdS/CFT correspondence \cite{Maldacena:1997re}.
In the context of emergent gravity, however, this observation indicates the {\it fluid/gravity equivalence} \cite{Mei:2022ksw,Mei:2022kcf}:
\begin{itemize}
\item The existence of our cosmic space (the kind of cosmic space that is familiar to us) is due to the existence of some hidden fluid, and the properties of spacetime and gravity are determined by the properties of the hidden fluid and how it interacts with matter.
\end{itemize}
The hidden fluid is expected to exist on some more fundamental supporting structure, which we call the {\it dry vacuum}.
Very much like a drop of water in the empty space, the hidden fluid is expected to occupy only a finite region of the dry vacuum.
For the model considered in \cite{Mei:2022ksw,Mei:2022kcf}, the expansion of the universe is correlated to the decrease in the density of the hidden fluid, and the universe would be infinitely expanded if the density of the hidden fluid goes to zero.
From the perspective of the dry vacuum, this indicates that the region without the hidden fluid is not traversable to everything that perceives a common expansion of our current universe.
As a result, our cosmic space is confined to the region of the dry vacuum permeated by the hidden fluid, leading to the \ac{BCS}.
To help visualize the picture, one may use the fish tank as an analogy: the hidden fluid is like the water in the fish tank, everything that propagates in our cosmic space is like fish in the fish tank, and that one is not able to get beyond the \ac{CB} is like fish cannot get away from the water.


Of course, the above motivation is helpful but is not crucial to the experimental search of the \ac{CB}.
Since there is no credible theory that can make convincing predictions on the global structure of the cosmic space, one may take the pragmatic point of view that, the presence (or not) of a \ac{CB} is something that can only be answered with experimental observations.

\begin{figure}
\centering\includegraphics[width=\textwidth]{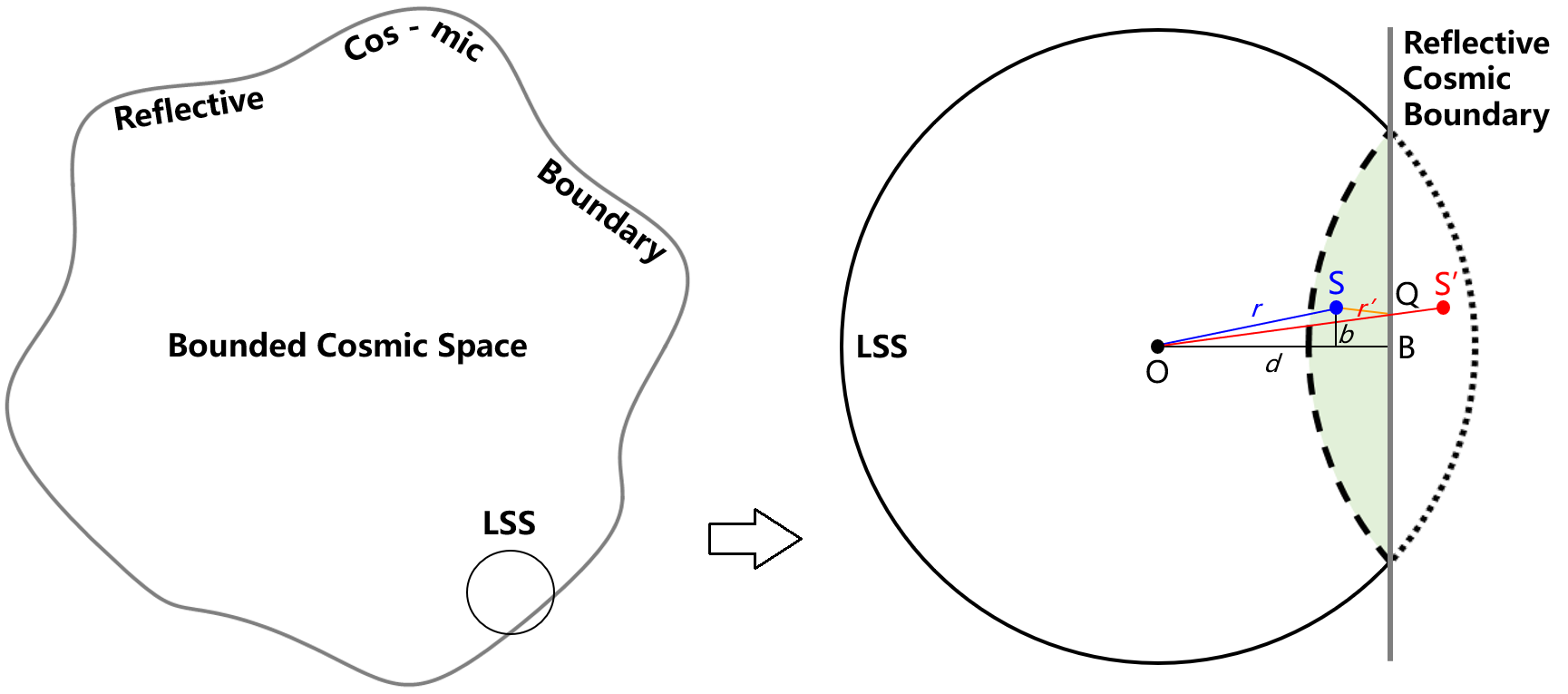}
\caption{(Left) A possible relation between LSS and BCS.
(Right) A zoom-in on the LSS.
All objects in the figures are assumed to have fixed comoving coordinates.
S is an astrophysical source and S' is its image in the \ac{CB}.
The path S'O intersects the \ac{CB} at the point Q.
OB is normal to the \ac{CB}. $r$, $r'$ and $d$ are the comoving radial coordinates of S, S' and B, respectively.
$b$ is the distance of S (and S') to the line OB. }
\label{fig:CB}
\end{figure}

For the properties of the \ac{CB}, there is no reason to believe that the size of the \ac{BCS} is similar to that of the \ac{LSS} or that the shape of the \ac{BCS} is simple and symmetric.
For example, the \ac{BCS} can be much larger than the \ac{LSS} and its shape can be distorted.
Of course, in order for the \ac{CB} to be relevant for observations, we need to focus on the possibility that the \ac{LSS} happens to be located to a side of the \ac{BCS} and intersects with part of the \ac{CB}, as illustrated in Fig. \ref{fig:CB} (Left).
To keep the discussion simple, we will assume that the portion of the \ac{CB} that is inside the \ac{LSS} is flat.
Fig. \ref{fig:CB} (Right) is a zoom-in on the \ac{LSS}.
The plane of the figure goes through the observer O in the solar system and is perpendicular to the \ac{CB}.
The \ac{CB} cuts off part of the \ac{LSS} and everything to the right of the \ac{CB} in fact does not exist.
The dashed curve is at the mirror position of the dotted curve.
If the \ac{CB} is fully absorptive, then there would be a dark patch on the \ac{CMB} sky.
Since this has not been observed, we assume that the \ac{CB} is fully reflective for both light and \acp{GW}.\footnote{It is natural to ask what happens if a star or a black hole hits the \ac{CB} and if the collision can produce detectable \ac{GW} signals. We leave such questions for future study.}
In this case, the dashed curve replaces the dotted curve to become the source of part of the \ac{CMB} photons, and the observer still has a complete \ac{CMB} map.

How does the \ac{CB} evolve as the universe expands?
This will depend on the detailed properties of the hidden fluid and the dry vacuum.
Here we focus on one particular possibility:
\begin{itemize}
\item Firstly, since the properties of gravity and spacetime are determined by the properties of the hidden fluid, variations in the properties of the hidden fluid could be fully responsible for the observed expansion of the universe (see \cite{Mei:2022ksw,Mei:2022kcf} for an explicit example).
\item Secondly, from the perspective of the dry vacuum, it is possible that the size and shape of the volume occupied by the hidden fluid are fixed.
\end{itemize}
In the fish tank analogy, the above correspond to saying that the expansion of the universe observed by the ``fish" is totally caused by the change of the properties of ``water", while the size and shape of the ``fish tank" stay fixed.
Motivated by this picture, we will assume that the \ac{CB} has fixed comoving coordinates, which do not change despite the expansion of the universe.

For the calculations of this paper, we will assume that the interior of the \ac{BCS} is described by the flat \ac{FLRW} metric, $ds^2=-dt^2+a(t)^2(dr^2+r^2d\Omega_2)\,$, for which the origin of the spatial coordinates is located at the observer O. The point B in Fig. \ref{fig:CB} is placed on the $\hat{x}$-axis, with the comoving Cartesian coordinates: $x_B=d\,$, $y_B=z_B=0$. The source S is located at: $x_S=\sqrt{r^2-b^2}\,$, $y_S=b\,$, $z_S=0\,$. The comoving distance and luminosity distance of an object at redshift $z$ are determined through $D_z=\int_0^z\frac{dz'}{H(z')}$ and $D_L(z)=(1+z)D_z$, respectively, where the Hubble parameter is
\bea H(z)=H_0\Big[\Omega_\Lambda +\Omega_M(1+z)^3 +(\Omega_\gamma+\Omega_\nu) (1+z)^4\Big]^{1/2}\,.\eea
The cosmological parameters are \cite{Lahav:2024npe}: $H_0=100 h$ km/s/Mpc with $h\approx0.677$, $\Omega_\Lambda\approx0.689$, $\Omega_m\approx0.311$,  $\Omega_\gamma \approx2.47\times10^{-5}h^{-2}$, and $\Omega_\nu \approx (\sum m_\nu) (93.12~{\rm eV})^{-1}h^{-2}$.
There is uncertainty in the neutrino masses, but which are most relevant only in the radiation dominant era.
We assume $\sum m_\nu=0.1$ eV in this paper \cite{Planck:2018vyg}.

\section{Detection scheme}
\label{sec:scheme}

Searching for a reflective \ac{CB} at the high redshift is not easy.
Firstly, such surface will have little detectable effect on the \ac{CMB} due to the homogeneity assumption about the universe, which indicates that the temperature variation along the dashed curve in Fig. \ref{fig:CB} is statistically indistinguishable from that along the dotted curve had it not been cutoff by the \ac{CB}.
The sign of the \ac{CMB} polarization \cite{Hu:1997hv} can be flipped by the reflective \ac{CB}.
But this also appears difficult to observe.
Secondly, one also cannot rely on parity related characters of individual astrophysical objects to tell if there is a reflective surface, because no universally parity violating character is known for astrophysical objects.
So to tell if there is a reflective surface, one must find a way to compare a source to its image caused by the \ac{CB}.

\begin{figure}
\centering\includegraphics[width=0.7\textwidth]{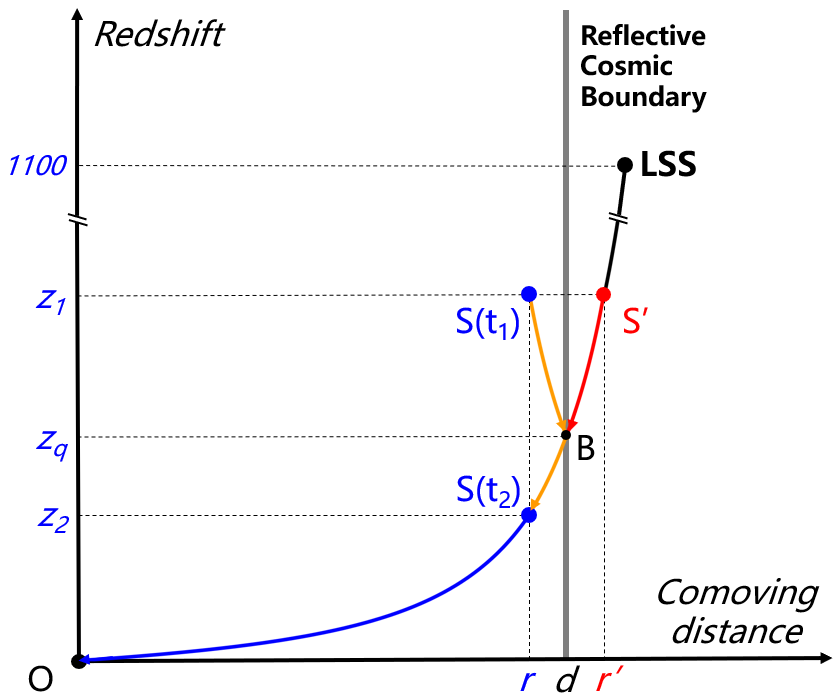}
\caption{The scheme to detect the \ac{CB}. The figure has been drawn with $b=0$ for illustration purpose.}
\label{fig:scheme}
\end{figure}

A possible detection scheme is illustrated in Fig. \ref{fig:scheme}, in which the past light cone of the observer along the spatial direction O$\rightarrow$B (corresponding to $b=0$ in Fig. \ref{fig:CB}) is shown.
All events/images that can be detected by the observer must lay on the light cone, which is represented by the curve LSS $\rightarrow$ S' $\rightarrow$ B $\rightarrow$ S($t_2$) $\rightarrow$ O in the figure.
The physical processes involved in the detection scheme are the following:
\begin{itemize}
\item The source S emits a signal at the time $t_1$, corresponding to the event S($t_1$);
\item The signal from S($t_1$) firstly travels towards the \ac{CB}, gets reflected, and then travels toward the observer. For the observer, this is equivalent to having a signal coming from the image S';
\item The source S emits a second signal at $t_2$, corresponding to the event S($t_2$);
\item The signals from S($t_1$) and S($t_2$) travel toward the observer concurrently, and both are eventually detected by a detector at the location of the observer.
\end{itemize}
For later convenience, we will refer to S($t_1$) as the first source event, S($t_2$) the second source event, S' the image event, S($t_1$) and S($t_2$) the source events, and S' and S($t_2$) the coincident events.

In this detection scheme, the distance of the \ac{CB} that can be explored is bounded by the largest possible distance of the image event.
Such bound can either come from the technology capability of the detectors or from the availability of suitable sources.
If the \ac{CB} is close enough, one may use an electromagnetic telescope to search for it.
In this case, the source events can be two snapshots of a same galaxy.
Currently, the most distant galaxy known is JADES-GS-z14-0, which has been detected by JWST at the redshift $z\approx14.3$ \cite{2024arXiv240518485C}.
This represents the highest redshift (for individual astrophysical objects) that one can explore with a current electromagnetic telescope.

For higher redshifts, one can use \acp{GW}.
In this case, the source events can be two merger events from the growth history of a \ac{MBH}.
\acp{MBH} can start from the remnants of Pop III stars, \acp{SBH}, massive seeds resulted from very massive stars, \acp{PBH},  and so on, and grow their masses through accretion and mergers \cite{LISA:2022yao}.
Among all the possible \ac{MBH} seeds, \acp{PBH} are particularly interesting because they might have been produced during the inflation \cite{Ivanov:1994pa,Garcia-Bellido:1996mdl,Ivanov:1997ia}.
If they can be used to detect the \ac{CB}, then one can reach redshifts higher than the formation of the first stars and galaxies.
Unfortunately, it has been shown that the chance for a \ac{PBH} to have second-generation merger is very low \cite{DeLuca:2020bjf}.
So we will focus on \acp{MBH} of non-\ac{PBH} origin in this paper.
This will limit the redshift range of the \ac{CB} that can be explored.
For example, if we fix the redshift of S' to be at $z=20$, then we can only explore the \ac{CB} with a redshift no higher than $z=20$.

Below we study what \ac{GW} sources can be used to detect the \ac{CB} and what will be needed to confirm the existence of the \ac{CB}.
For all the calculations, unless otherwise specified, we will use the IMRPhenomXHM waveform \cite{Garcia-Quiros:2020qpx}, and focus on black hole binaries with the major mass (source frame) $m_1=1\times10^5\mSun$, mass ratio $q=m_1/m_2=1.2$, source inclination angle $\iota=0.9$ rad, dimensionless spins $s_1=0.4$ and $s_2=0.2$ (both are assumed to be aligned with the orbital angular momentum), and will assume an observation time $T_{\rm obs}=$ 1 month.

\subsection{GW detectors and required source parameters}

Several space-based \ac{GW} detectors are being planned for the 2030s \cite{Gong:2021gvw}.
In this work, we use TianQin \cite{Luo:2015ght,TianQin:2020hid} and LISA \cite{LISA:2017pwj} as examples.

TianQin will be consisted of three Earth orbiting satellites, forming a nearly normal triangle constellation with each side measuring about $L_{\rm TQ}\approx \sqrt{3}\times10^8$ m.
The sensitivity of TianQin is given by \cite{Luo:2015ght,TianQin:2020hid}:
\bea S_n^{\rm TQ}(f) = \frac{10}{3 L_{TQ}^2} \left[S_x^{\rm TQ} +\frac{4 S_a^{\rm TQ}}{(2 \pi f)^4} \Big(1 + \frac{10^{-4}{\rm Hz}}{f}\Big) \right ] \times  \left[ 1 + 0.6 {\left(\frac{f}{f_*^{\rm TQ}} \right)}^2 \right ],
\eea
where $(S_x^{\rm TQ})^{1/2}=1\times 10^{-12} {\rm m}/{\rm Hz}^{1/2}$ is the one-way displacement measurement noise, $(S_a^{\rm TQ})^{1/2} =1 \times 10^{-15}$ m/s$^2$/Hz$^{1/2}$ is the residual acceleration of a test mass along the sensitive axis, and $f^{\rm TQ}_* = 1/(2 \pi L_{\rm TQ})\approx 0.28$ Hz is the transfer frequency of TianQin.

LISA will be consisted of three spacecraft forming a nearly normal triangle constellation with each side measuring about $L_{\rm LISA}\approx 2.5\times10^9$ m.
The center of the LISA constellation is on the same orbit as the Earth and is at a distance of about 50 million kilometers from the Earth.
The plane of the LISA constellation is at about 60$^\circ$ with respect to the ecliptic plane.
Apart from its annual orbital motion around the Sun, the LISA constellation also has an annual cartwheel motion.
The sensitivity of LISA is given by \cite{Robson:2018ifk}:
\bea S_n^{\rm LISA}(f) &=& \frac{10}{3 L_{\rm LISA}^2}
\left\{S_x^{\rm LISA}(f) + \frac{S_a^{\rm LISA}(f)}{{(2\pi f)}^4}
\left[1 + \cos^2\left( \frac{f}{f_*^{\rm LISA}} \right)\right]\right\}
\left[ 1 + \frac{6}{10} {\left( \frac{f}{f_*^{\rm LISA}} \right)}^2 \right]\,,\eea
where $f_*^{\rm LISA} = 1 / (2 \pi L_{\rm LISA})\approx0.02$ Hz is the transfer frequency of LISA, and
\bea S_x^{\rm LISA}(f)&=& S_x^{\rm LISA}\left[1+{\left( \frac{2 \times 10^{-3} {\rm Hz}}{f} \right)}^4\right]\,,\nn\\
S_a^{\rm LISA}(f)&=&S_a^{\rm LISA}\left[1 + {\left( \frac{4 \times 10^{-4}{\rm Hz}}{f}\right)}^2\right]\times\left[1 + {\left( \frac{f}{8\times 10^{-3}{\rm Hz}}\right)}^{4}\right]\,,\eea
where $(S_x^{\rm LISA})^{1/2}=1.5\times 10^{-11}$ m/Hz$^{1/2}$ and $(S_a^{\rm LISA})^{1/2}=3\times10^{-15}$ m/s$^{2}$/Hz$^{1/2}$.

Both TianQin and LISA have the goal to detect \acp{GW} in the frequency range $10^{-4}$ Hz $\sim$ 1 Hz.
But LISA has longer arm-length and can potentially cover a wider frequency range, i.e., $2\times 10^{-5}$ Hz $\sim$ 1 Hz \cite{LISA:2017pwj}.
It is known that the detection capability can be significantly improved if the space-based \ac{GW} detectors can form a network (see, e.g., \cite{Ruan:2020smc,Gong:2021gvw}).
This is also true for the TianQin+LISA network.
A detailed study of how TianQin+LISA can improve over each individual detectors can be found in \cite{Torres-Orjuela:2023hfd}.

\begin{figure}[!htbp]
\centering
\subfigure{\includegraphics[width=0.48\textwidth]{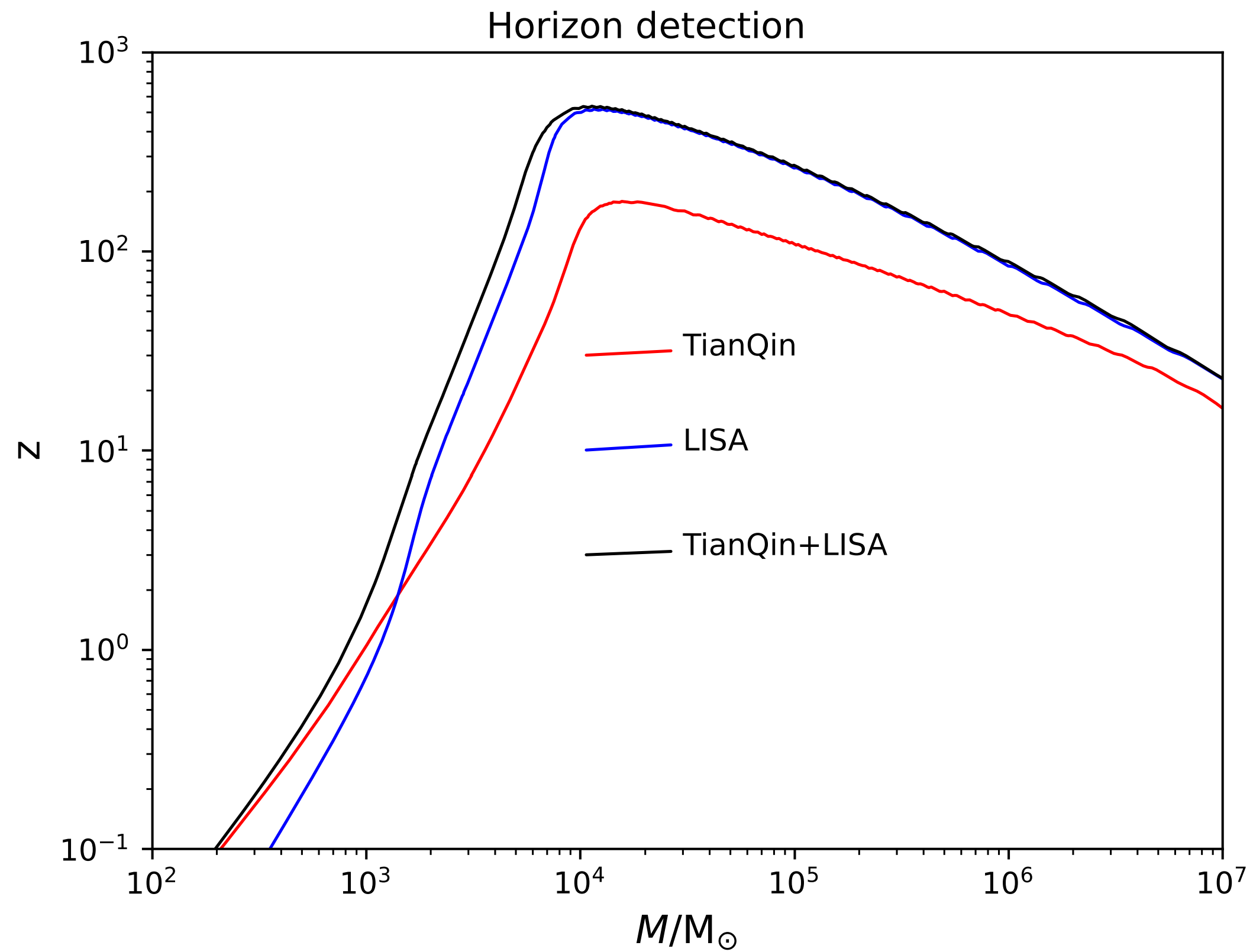}}
\subfigure{\includegraphics[width=0.48\textwidth]{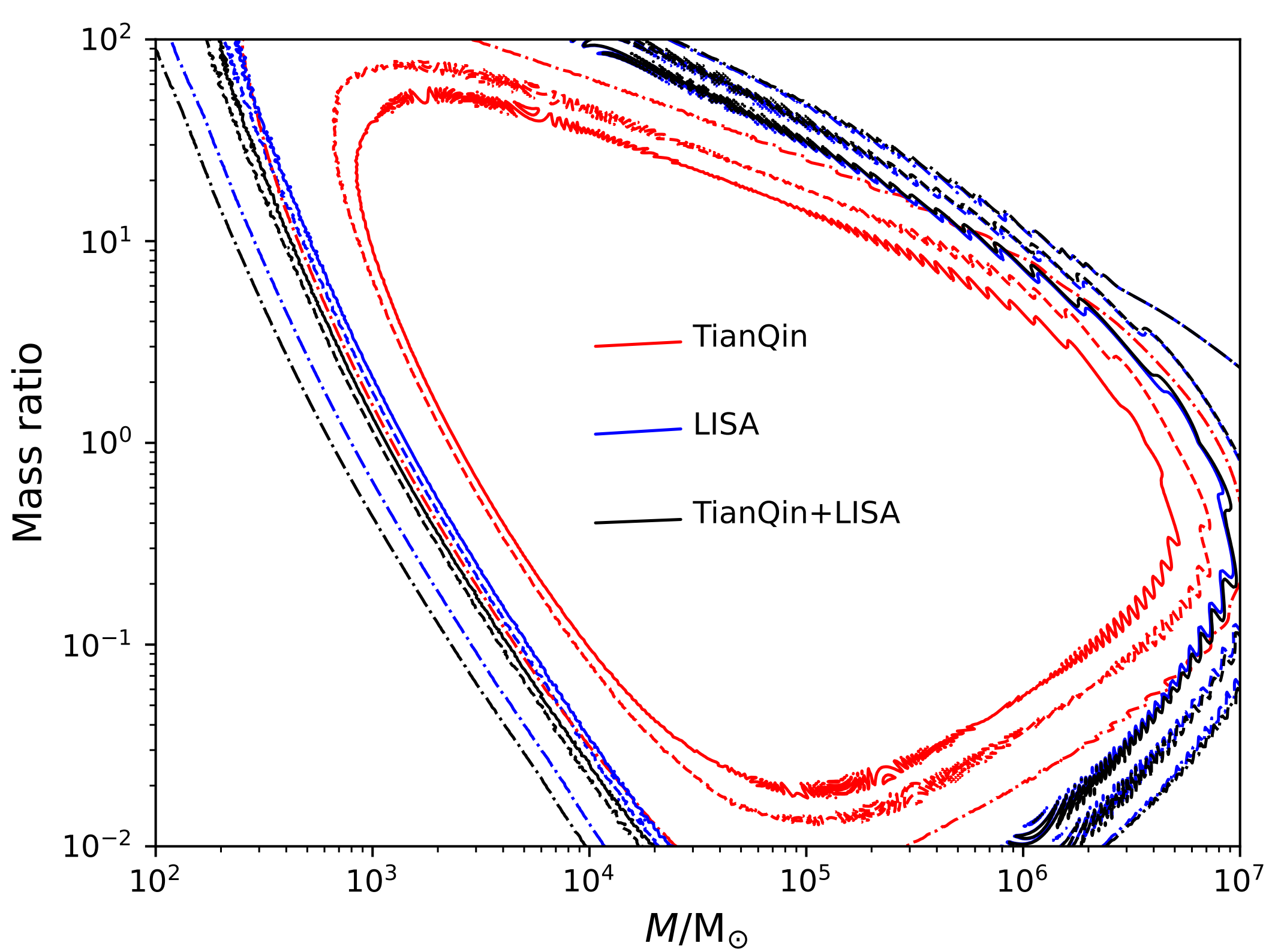}}
\caption{(Left) The detection horizon of TianQin, LISA and TianQin+LISA. (Right) Allowed mass range for candidate GW sources, for S' at $z=20$ (solid), $S(t_2)$ at $z_2=16.2$, corresponding to $z_q=18$ (dashed), and $S(t_2)$ at $z_2=11.6$, corresponding to $z_q=15$ (dot-dashed).}
\label{fig:horizon}
\end{figure}

Fig. \ref{fig:horizon} illustrates the detection horizon of TianQin, LISA and TianQin+LISA, and  the allowed mass range for possible candidate \ac{GW} sources at different redshifts, assuming the detection threshold SNR $=8$. One can see that both TianQin and LISA can reach $z>15$ for a wide range of source masses.

\subsection{Identifying the coincident events}

In order to find the \ac{CB}, one must firstly detect the coincident \ac{GW} events.
To identify the coincident events, one can check the following:
\begin{itemize}
\item \underline{{\it Angular separation}} As indicated in Fig. \ref{fig:CB}, the angular separation between the coincident events, $\angle$SOS', must lay within a certain range determined by the distance of the \ac{CB}, represented by the redshift $z_q$ of the point B, and the orientation of the \ac{CB}, represented by the angle $\angle$S'OB.

\item \underline{{\it Mass}}
For a pair of coincident \ac{GW} events, one component of the second source event S($t_2$) must be inherited from the remnant black hole of the first source event S($t_1$).
The remnant black hole of S($t_1$) might have experienced some accretion and merger in between S($t_1$) and S($t_2$).
So for a pair of coincident events, S($t_2$) must have one component no less massive than the remnant of S($t_1$) and equivalently of S'.

\item \underline{{\it Spin}}
Due to the reflection by the \ac{CB}, a spin parallel to the \ac{CB} will have its sign flipped, while that perpendicular to the \ac{CB} will stay the same.
This will lead to a predictable relation between the spin of the remnant of S' and that of one of the components of S($t_2$), given that the location of \ac{CB} is known and that the spin of the remnant of S($t_1$) has not been significantly altered by accretion or merger in between S($t_1$) and S($t_2$).
\end{itemize}

\begin{figure}
\centering\includegraphics[width=\textwidth]{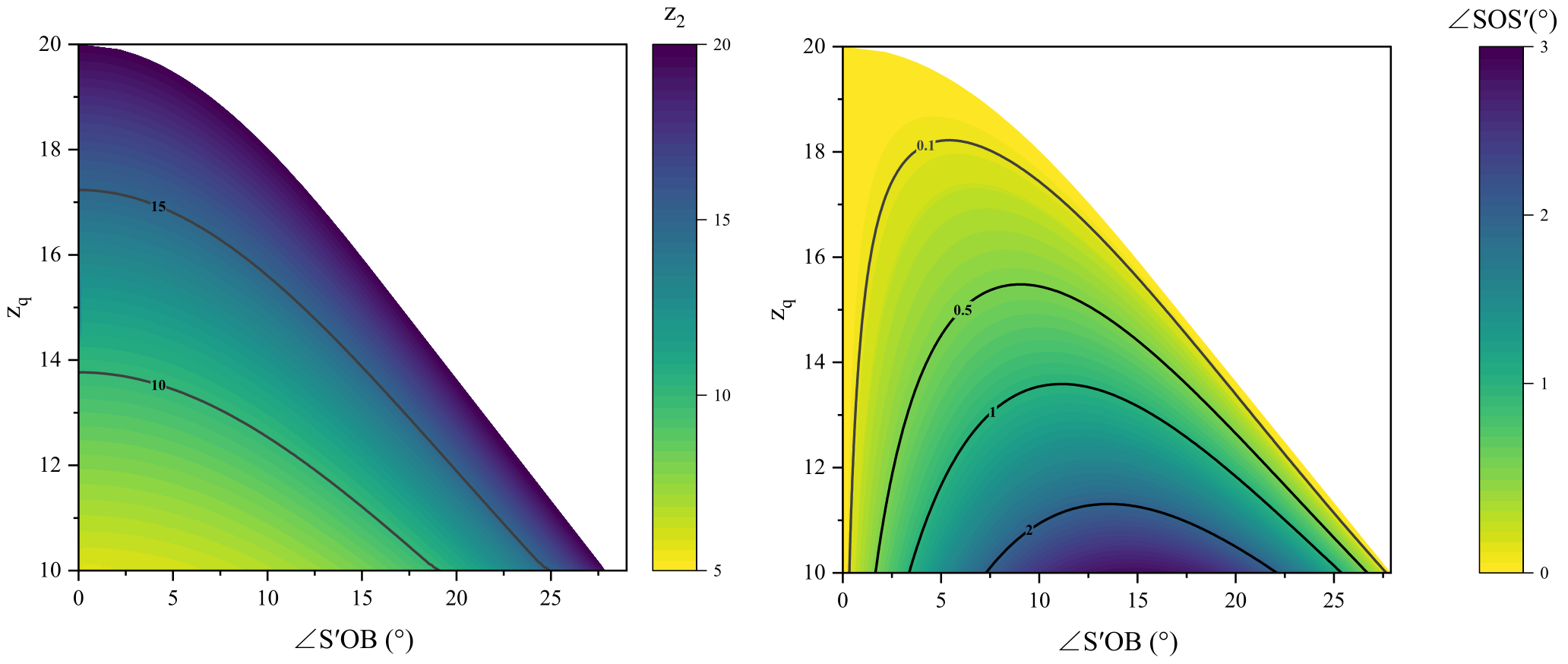}
\caption{The dependence of $z_2$ and $\angle$SOS' on $\angle$S'OB and $z_q$.}
\label{fig:angle}
\end{figure}

Fig. \ref{fig:angle} illustrates the dependence of $z_2$ and $\angle$SOS' on $\angle$S'OB and $z_q$.
In the figure, the image event S' is fixed at the redshift $z_1=20$, and $z_q$ is varied from 10 to 20.
Correspondingly, $z_2$ can vary from about 5 to 20, which also depends on $\angle$S'OB.
One can see that for $z_q>15$, one always has approximately $\angle$SOS'$<0.5^\circ$.
Such a small angular separation between the coincident events is helpful in that it can significantly shorten the list of potential candidates.

\begin{figure}
\centering
\subfigure{\includegraphics[width=0.48\textwidth]{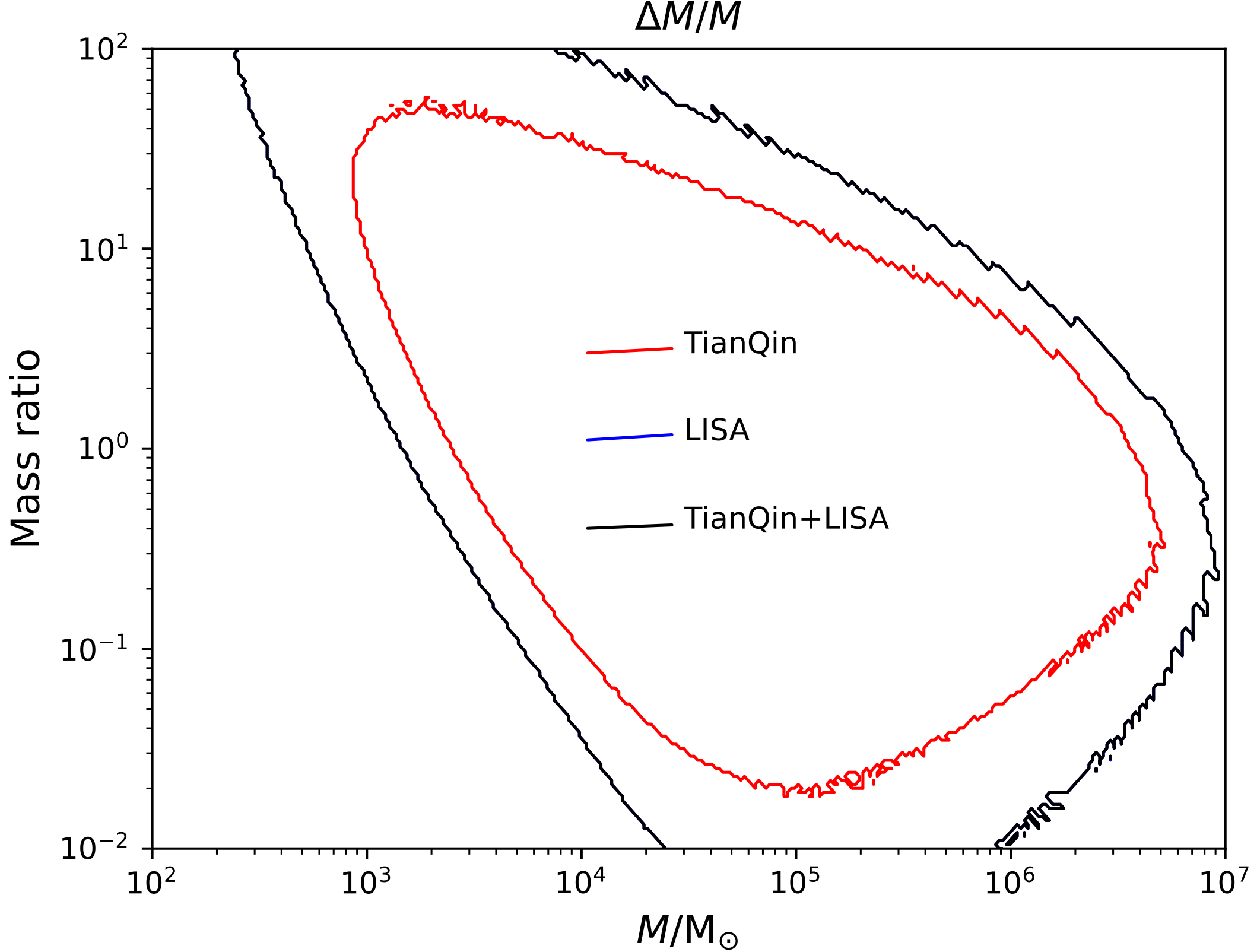}}
\subfigure{\includegraphics[width=0.48\textwidth]{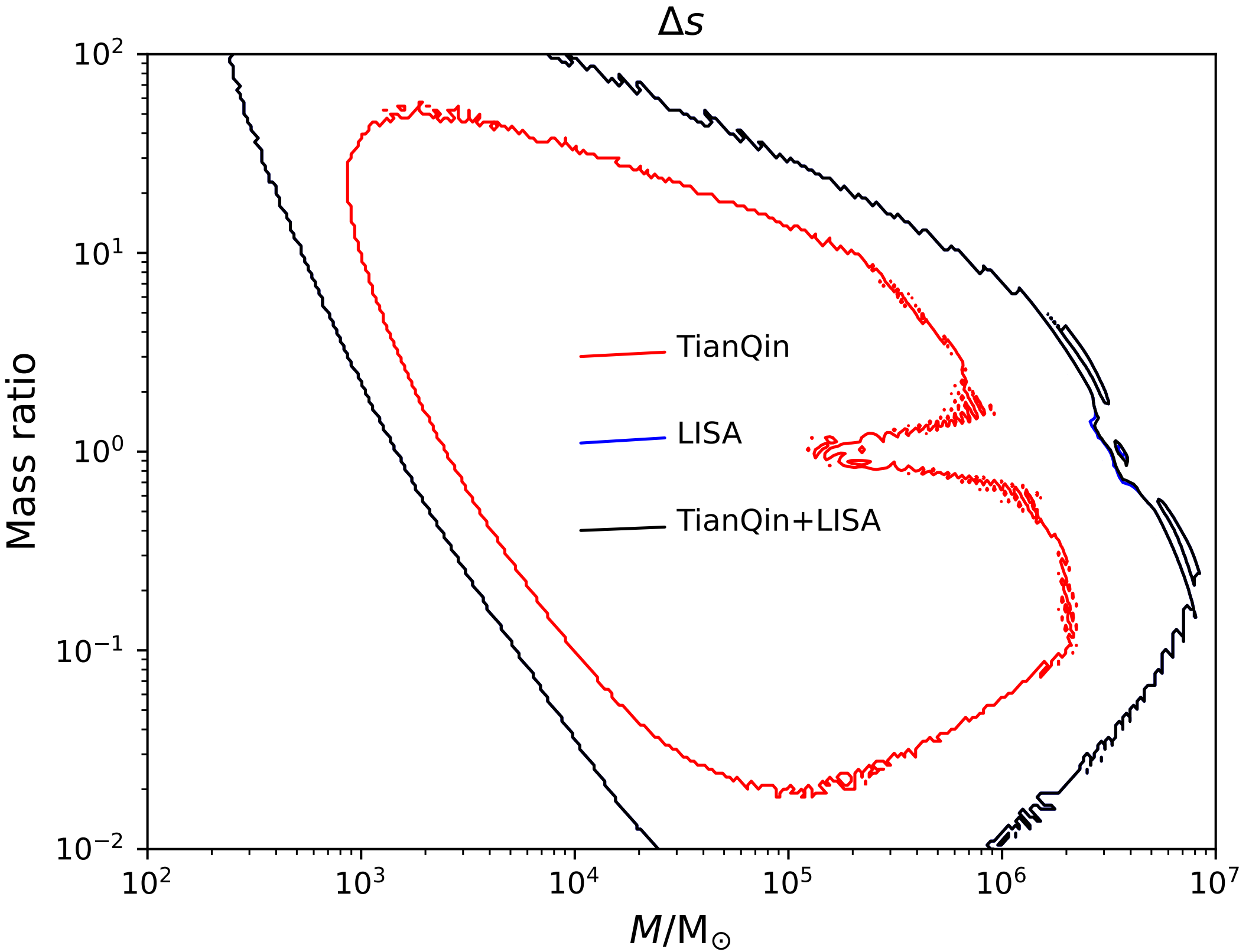}}
\centering\includegraphics[width=0.48\textwidth]{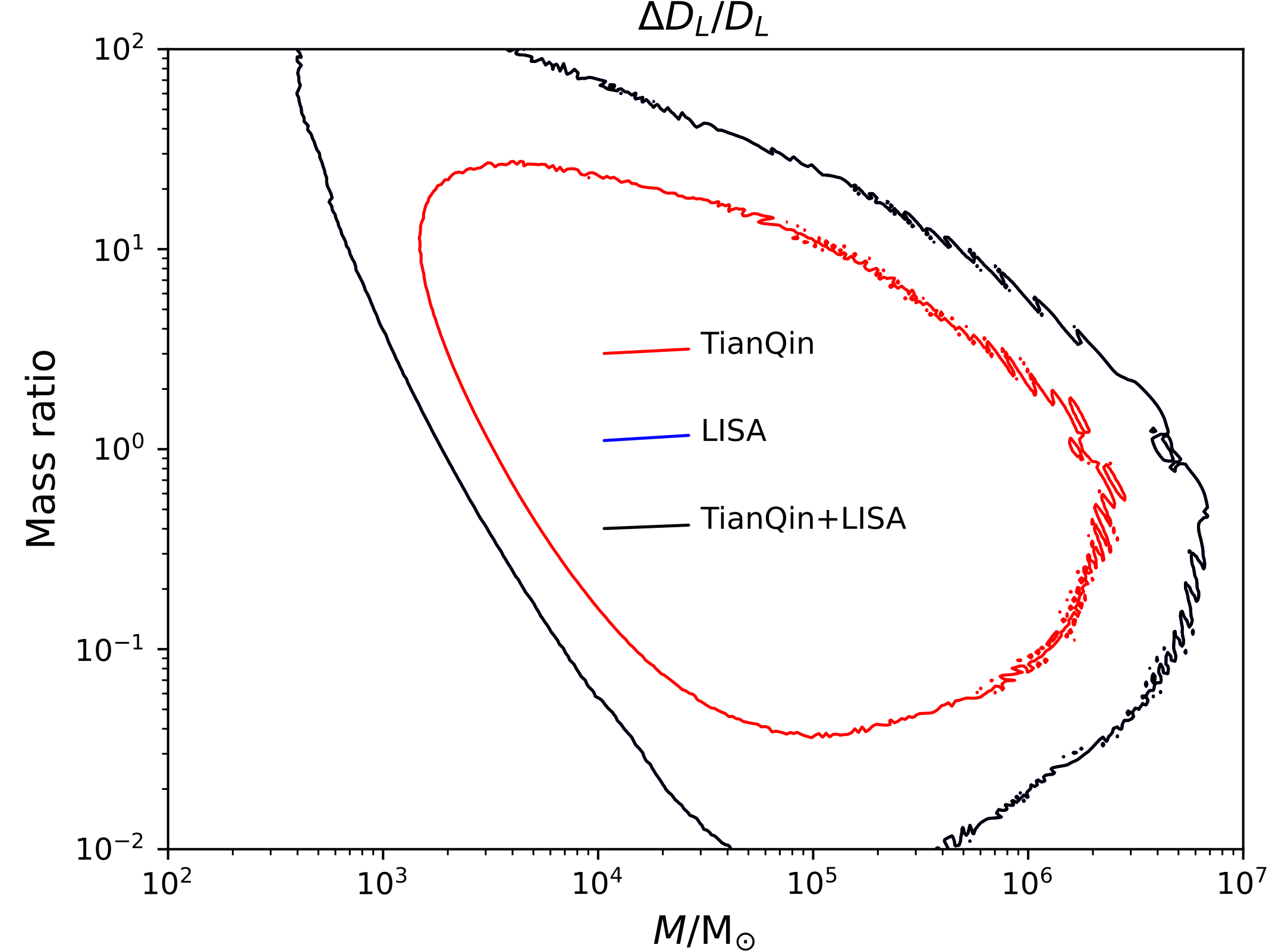}
\caption{The expected precision for measuring mass, spin and luminosity distance for a source located at $z=20$. Regions inside the contours are for $\Delta M/M<10\%$ (top left), $\Delta s<0.1$ (top right) and $\Delta D_L/D_L<10\%$ (buttom). Presented in the sequence: (TianQin, LISA, TianQin+LISA), the highest precisions in each plot are: $\Delta M/M=(2.7, 1.4, 1.2)\times10^{-5}$, $\Delta s=(2.5, 1.2, 1.1)\times10^{-4}$ and $\Delta D_L/D_L=(12.4, 4.3, 4.1)\times10^{-3}$.}
\label{fig:mass}
\end{figure}

Once the extrinsic parameters match, one can then look at the intrinsic parameters.
Fig. \ref{fig:mass} (top left and right) illustrates the expected measurement precision for the mass $M$ and dimensionless spin $s$ of the component of $S(t_2)$ that is inherited from the remnant black hole of $S(t_1)$.
For a worst case scenario, we focus on $z_2\approx z_q\approx z_1=20$.
The remnant black hole will become a component in the merger event of $S(t_2)$.
For these parameters, LISA dominates the contribution to the TianQin+LISA network.
One can see that, even for sources located at redshift as high as $z=20$, one still has $\Delta M/M<10\%$ and $\Delta s<0.1$ for a wide range of source masses.
The best precisions can reach $\Delta M/M\sim \cO(10^{-5})$ and $\Delta s\sim\cO(10^{-4})$.

\subsection{Necessary conditions to confirm the CB}

With the present detection scheme, it is difficult to confirm a pair of coincident events to the 100\% level, even when all the extrinsic and the intrinsic parameters are as predicted.
In order to confirm the presence of the \ac{CB} to a high confidence level, it is necessary to have multiple pairs of coincident events, with all of which predicting the same location for the \ac{CB}.
This will be challenging not only in terms of the number of events that need to be detected, but also in terms of the precision that one needs to locate the sources.

A reliable assessment of the chance to detect multiple pairs of coincident events is difficult and is beyond the scope of this paper.
Here we only want to remark that, given the weight that the possible presence of the \ac{CB} bears on our entire view of the universe, it is always worthwhile searching for potential signatures of it, no matter how small the chance is.

For the precision to determine the space location of the sources, Fig. \ref{fig:mass} (bottom) illustrates the expected precision for measuring the luminosity distance of $S(t_2)$ at $z_2\approx20$.
One can see that the luminosity distance can be determined to better than $\Delta D_L/D_L = 10\%$ for a large range of source masses.
The best precisions can reach $\Delta D_L/D_L\sim\cO(10^{-3})$.

\begin{figure}
\centering
\subfigure{\includegraphics[width=0.8\textwidth]{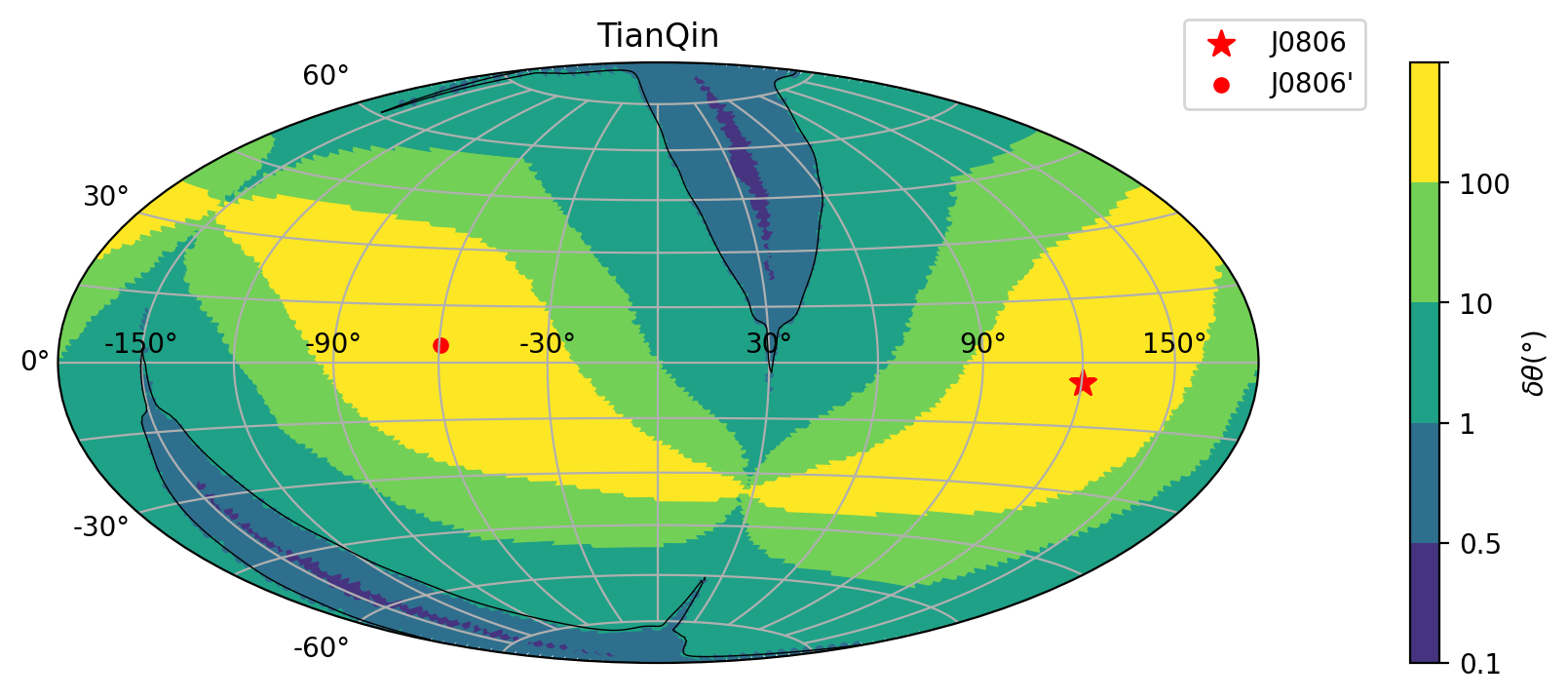}}
\subfigure{\includegraphics[width=0.8\textwidth]{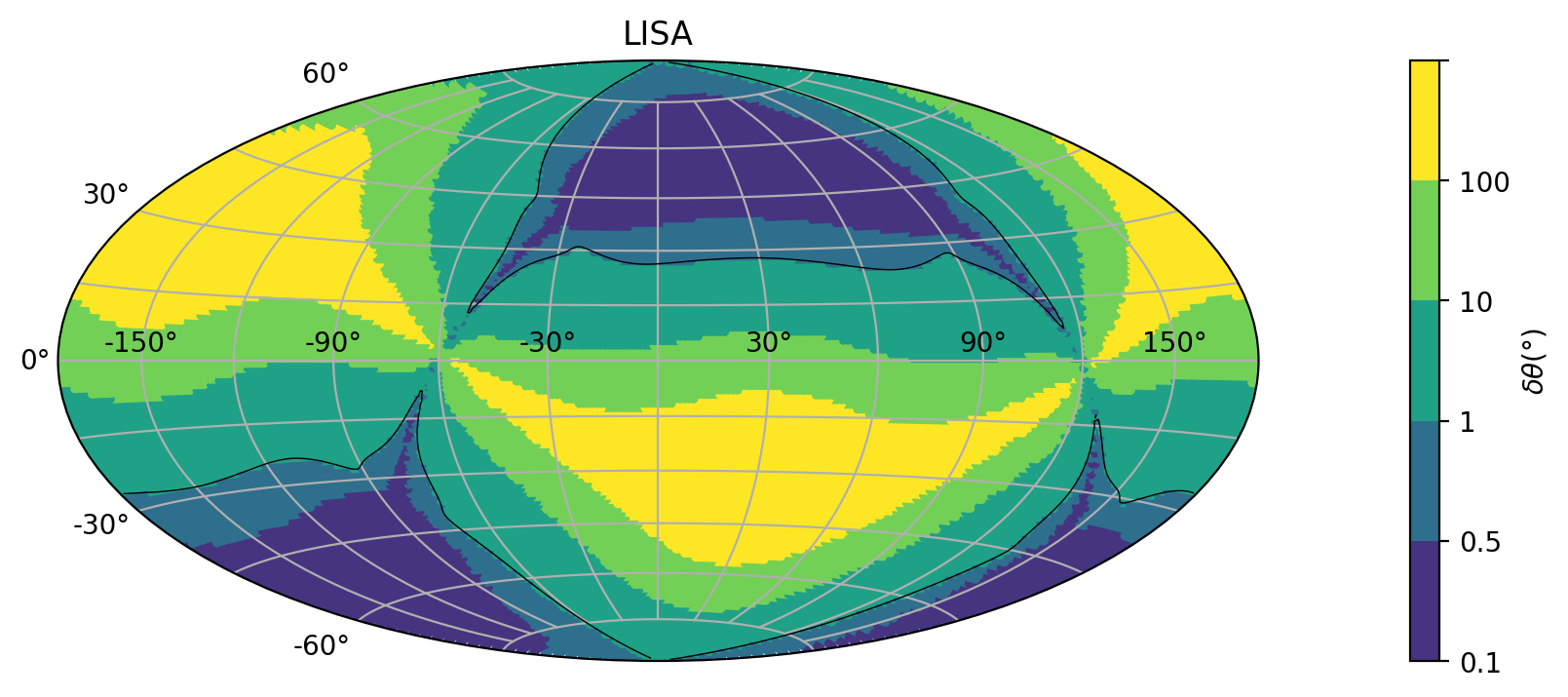}}
\subfigure{\includegraphics[width=0.8\textwidth]{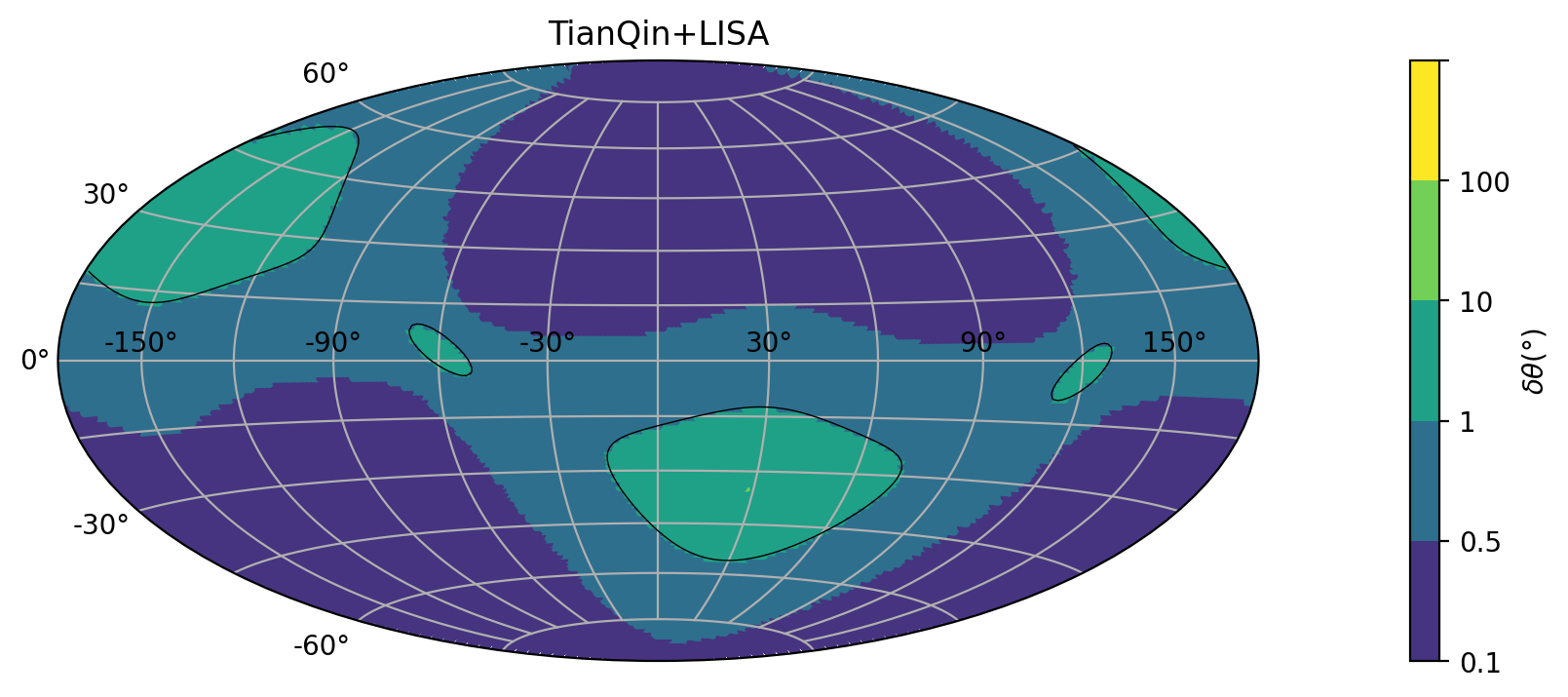}}
\caption{The expected precision of angular resolution for a source at $z=20$ with TianQin, LISA and TianQin+LISA. J0806' indicates the opposite direction of J0806. The orbital angular momenta of all the sources are fixed at the direction $\theta_L=\phi_L=0.6$ rad. The exact shape of the contours will change significantly if one modifies the source parameters, but the contrast between the capability of different detectors and detector network is roughly the same.}
\label{fig:omega}
\end{figure}

From Fig. \ref{fig:angle}, one can see that $\angle$SOS' does not depend on $\angle$S'OB very sensitively.
This means that one will need to measure $\angle$SOS' very precisely to determine the orientation of \ac{CB}.
For example, for $z_q>15$, one needs to measure $\angle$SOS' to better than about $0.5^\circ$.
Fig. \ref{fig:omega} illustrates the expected precision of the angular resolution, $\delta\theta=\sqrt{\Delta\Omega/\pi}$, for a source located at $z=20$.
In the plots, the exact shape of the contours significantly depends on the choice of the source parameters, but the contrast between the capability of different detector configurations is roughly the same.
One can see that the TianQin+LISA network can significantly increase the region of the sky in which a source can be localized to better than $\delta\theta\sim\cO(0.5^\circ)$.
So a detector network like TianQin+LISA is essential in help increasing the chance to determine the location of the \ac{CB}, given that enough pairs of coincident events can be detected.

Fig. \ref{fig:omega} also shows that for $z_q>18$, the angular resolution is always worse than $0.1^\circ$ even with the TianQin+LISA network.
From Fig. \ref{fig:angle}, this means that one will not be able to determine the orientation of \ac{CB} in this case.
But the brighter side is, Fig. \ref{fig:angle} also indicates that one always has $\angle$S'OB $<10^\circ$ for $z_q>18$.

Finally, we remark that the above conclusions have been obtained by fixing the source at $z=20$.
The quantitative result can become different if it is located at other redshifts.

\section{Summary}
\label{sec:sum}

In this paper, we have studied the potential of using \acp{GW} to detect the possible presence of a \ac{CB} at the high redshift.
The \ac{CB} is assumed to have fixed comoving coordinates, being reflective for both electromagnetic waves and \acp{GW}.
We focus on \acp{MBH} of non-\ac{PBH} origin and find that, with the future space-based \ac{GW} detectors like TianQin and LISA, a large variety of \acp{MBH} with masses roughly in the range $\cO(10^3\mSun)\sim \cO(10^6\mSun)$ can be used to detect the \ac{CB}.

To find the \ac{CB}, we have studied a detection scheme that relies on detecting two merger events from the growth history of a \ac{MBH}.
We find that, for a \ac{CB} located at the high redshift ($z_q>15$), the angular separation between a pair of coincident events is always smaller than about $0.5^\circ$.
The relative precision of mass and luminosity distance can be determined to better than the 10\% level, and the dimensionless spin can be determined to better than the 0.1 level.
Combined, these capabilities can provide strong evidence to identify possible pairs of coincident \ac{GW} events.

In order to confirm the presence of the \ac{CB} to a high confidence level, however, it is necessary to detect multiple pairs of coincident events.
In this case, a detector network like TianQin+LISA is crucial in help improving the chance to precisely determine the orientation of the \ac{CB}.
The chance for detecting a pair of coincident events is presumably low, and the chance to detect multiple pairs of coincident events is even lower.
So the possibility to prove or disprove the presence of the cosmic boundary largely depends on how likely one can detect multiple pairs of coincident gravitational wave events.
This is a key restriction to the detection scheme described in this paper.

Due to the importance of the possible nontrivial global structure of the cosmic space to our entire view of the universe, however, it is always worthwhile searching for potential signatures of it in the \ac{GW} data expected in the 2030s, even if the predicted chance is very small.

\section*{Acknowledgments}

This work has been supported in part by the Guangdong Major Project of Basic and Applied Basic Research (Grant No. 2019B030302001), the National Science Foundation of China (Grant No. 12261131504, 12405080), and the National Key Research and Development Program of China (Grant No. 2023YFC2206700).

\bibliographystyle{apsrev4-1}
\bibliography{main}
\end{document}